\documentclass[preprint,12pt]{elsarticle}

\usepackage{float}
\usepackage{graphicx,subfigure}
\usepackage{bm}% bold math
\usepackage{amsmath}
\usepackage{amssymb}
\usepackage{latexsym}
\usepackage{epsfig}
\usepackage{amsbsy}
\usepackage{array}
\usepackage{amssymb}
\usepackage{booktabs}
\usepackage{bbding}
\usepackage{multirow}
\usepackage{enumerate}
\usepackage[normalem]{ulem}
\usepackage[ruled,norelsize]{algorithm2e}
 \usepackage[misc]{ifsym}
\usepackage{graphicx}
\usepackage[mathlines]{lineno}
\usepackage{amssymb}
\biboptions{sort&compress}
\journal{Optics and Lasers in Engineering}

\begin{document}

\begin{frontmatter}

%% Title, authors and addresses

%% use the tnoteref command within \title for footnotes;
%% use the tnotetext command for theassociated footnote;
%% use the fnref command within \author or \address for footnotes;
%% use the fntext command for theassociated footnote;
%% use the corref command within \author for corresponding author footnotes;
%% use the cortext command for theassociated footnote;
%% use the ead command for the email address,
%% and the form \ead[url] for the home page:
%% \title{Title\tnoteref{label1}}
%% \tnotetext[label1]{}
%% \author{Name\corref{cor1}\fnref{label2}}
%% \ead{email address}
%% \ead[url]{home page}
%% \fntext[label2]{}
%% \cortext[cor1]{}
%% \address{Address\fnref{label3}}
%% \fntext[label3]{}

\title{Modified integral imaging reconstruction and encryption using an improved SR reconstruction algorithm}

%% use optional labels to link authors explicitly to addresses:
%% \author[label1,label2]{}
%% \address[label1]{}
%% \address[label2]{}

\author[author1]{Xiaowei Li}
\author[author1]{Ying Wang}
\author[author2]{Qiong-Hua Wang\corref{corr}}
\ead{qionghua@buaa.edu.cn}
\cortext[corr]{Corresponding author.}
\author[author3]{Yang Liu}
\author[author1]{Xin Zhou}

\address[author1]{School of Electronics and Information Engineering, Sichuan University, Chengdu 610065, China}
\address[author2]{School of Instrumentation Science and Opto-electronics Engineering, Beihang University, Beijing 100083, China}
\address[author3]{School of Computer and Information Technology, Liaoning Normal University, Dalian, 116082, China}

\begin{abstract}
%% Text of abstract
We propose a monospectral image encryption method in which the multispectral color image acquisition by using heterogeneous monospectral cameras. Because the captured monospectral elemental images (EIs) belongs to grayscale image, it is means that the captured EIs can be directly encrypted by the proposed encoding method. Subsequently, the linear cellular automata (CA) and hyperchaotic encoding algorithm are employed to encrypt the captured EIs. Different from previous methods, the proposed method can directly encrypt the multispectral color information rather than having to divide into three color channels (R, G and B), thereby, the proposed method can greatly reduce the encryption calculation.
\end{abstract}

\begin{keyword}
%% keywords here, in the form: keyword \sep keyword
Monospectral image encryption \sep Hyperchaotic system \sep Cellular automata \sep Color image encryption.
%% PACS codes here, in the form: \PACS code \sep code

%% MSC codes here, in the form: \MSC code \sep code
%% or \MSC[2008] code \sep code (2000 is the default)

\end{keyword}

\end{frontmatter}

%% \linenumbers

%% main text
\section{Introduction}

\label{sec:1}
Recently, the multimedia security such as color images is drawing more and more attention because of its importance in various applications such as e-education and military that lead to the importance of real-time and sufficiently secure and robust image encryption methods \cite{liu2001,Wang2018,Zhang2018,Li201808,hua2015,Qin2018,bao2015}. The color image encryption algorithm is quite different from those of textual or binary data, and that is because of its special features of color images such as large size, high redundancy (three color channels) and large computation \cite{Wang2016,Chen2017,Abuturab2017,Sui2018,Suio2017,Sui2017,Joshi2008,Liu2001,Chen201401,Alfalou2009,Situ2005,Peng2006,Wang201601,Jiao2017,Liu2010,Li2018}.

Chaos for image security have triggered much interest, such as topological transitivity and random-like behaviors. Considering the features and the performance of the chaotic algorithms, Britain mathematician Matthes \cite{Matthews1989} firstly introduced chaotic theory for image encryption. After that, lots of image encryption methods based on chaos have been presented \cite{wangx2013,Hizanidis2010,Gao2008}.

It is very vital for an effective encryption method to be sensitive to the encryption keys, thus the key space of the encryption method should be large enough to resist brute-force attacks \cite{lic2017,lic2018,lix2015,zhou2016,lix2016}. Recently, cellular automata (CA) based image encryption methods have been presented because of the capacity of parallel calculation and high security. CA are dynamical systems on finite state sets \cite{Bidlo2016,Tsompanas2015,lix2017}. Pseudorandom number obtained by CA, which plays a very important role in image encryption. It is necessary to generate pseudorandom numbers in parallel owing to the emergence of massively parallel computation.

In the pickup process of integral imaging, the researchers utilize the Bayer color filter array (CFA) to capture 3D scene. Then 3D scene can be reconstructed by the computing integral imaging (CIIR) algorithm \cite{Latorre-Carmona2015,Javidi2017,Javidi2006,Andriani2013}. With the Bayer CFA, color image information can be efficiently acquired from single sensor camera. However, color crosstalk between each of color filters, which reduces the quality of the reconstructed 3D image \cite{Andriani2013}.

In this paper, a monospectral image encryption method is proposed. Different from the previous methods, in our work, the color image is first recoded into monospectral elemental images (EIs) by monospectral camera array. Because EIs belongs to the grayscale image, it can be directly encrypted by the proposed encoding algorithm, rather than having to divide into three color channels (R, G and B). Hence, it will greatly enhance the efficiency of color image encryption. Following that, we propose an improved super-resolution (SR) reconstruction technique to improve the image quality.

\section{Description of the monospectral image encryption algorithm}
\label{sec:2}

\begin{figure}
\centering
%\begin{minipage}[t]{\figwidth}
\centering
\includegraphics[width=\textwidth]{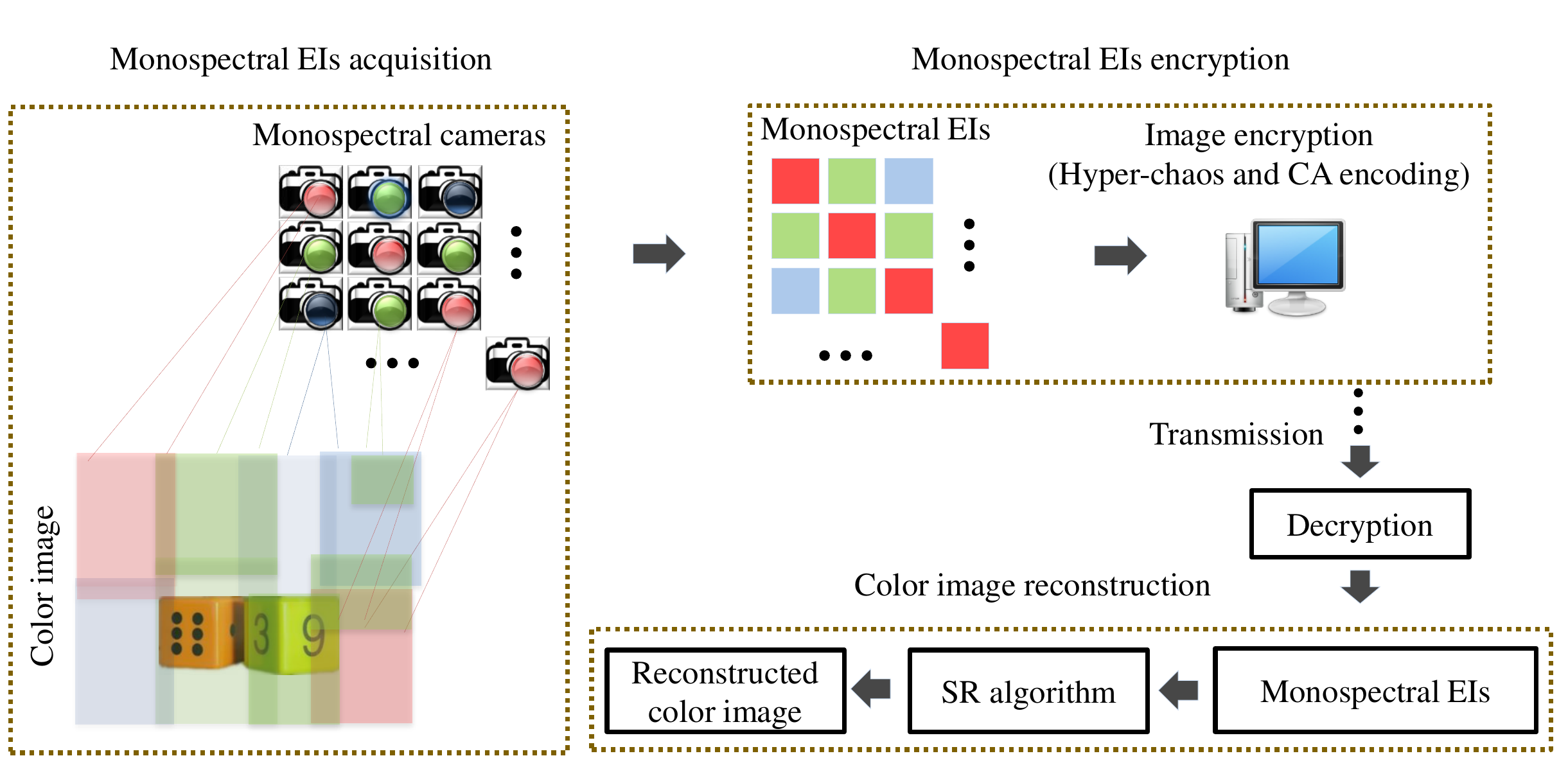}
%\end{minipage}
\caption{The implementation principle of the proposed color image encryption.}
\label{fig:en_sys}
\end{figure}

The encryption procedure of the proposed method is summarized in Fig.~\ref{fig:en_sys}. It can be divided into mainly three procedures: firstly, the color image is captured by heterogeneous monospectral cameras. Secondly, the captured monospectral EIs are encrypted by combining the use of the hyper-chaotic system and the CA encoding algorithm. Thirdly, the color image can be reconstructed by the inverse encoding process of encryption method and the improved SR image reconstruction method. The detailed description of the proposed color image encryption is presented below.

\subsection{Monospectral EIs captured by monospectral camera array}

To effectively reduce the color crosstalk, in our work, the monospectral camera is utilized to capture the color image. The left of Fig.~\ref{fig:en_sys} shows the monospectral EIs acquisition process through heterogeneous monospectral cameras. In this experiment, we do not use the Bayer color filter pattern of CCD camera; each camera of our proposed image capture method is isolated and only sensitive to just monochromatic color, red, green, or blue. The numbers of luminance sensors (green) of the monospectral camera array are more than the chrominance sensors (red, blue). The reason is that the human eyes are more sensitive to the luminance signals rather than the chrominance signals. For example, in a 4 $\times$ 4 monospectral camera array mode, eight or more sensors are sensitive to green spectrum for the increased image quality requirements. A sub-image array (SIA) is obtained through the monospectral camera array and each sub-image only possesses one of the spectral information (R, G, or B).

\subsection{Monospectral EIs encrypted by linear CA and hyper-chaotic system}
\label{sec:2.2}

\subsubsection{M-sequence generated by linear CA}
\label{subsubsection:2.21}

CA can offer significant benefit over existing algorithms \cite{Bidlo2016,Tsompanas2015}. In a one-dimensional (1D), two-state, three-site neighbourhood CA, where the next state of a cell is updated according to its neighborhood, the value of each cell is 0 or 1. The value of each cells can be generated by a specified rule. The value of next state is calculated by the Boolean function with three parameters.

\begin{equation}
s_i(t+1)=F(y_{i-1}(t),y_i(t),y_{i+1}(t)),
\end{equation}
where $s_{i}(t+1)$ denotes the value of cell $i$ at time $t+1$, $s_{i}(t)$  denotes the value of the cell $i$ at time $t$, $y_{i-1}(t)$ is the the left neighboring cell value at time $t$, $s_{i+1}(t)$ is the right neighboring cell value at time $t$, and $F()$ represents the Boolean function defining a specified rule.

According his theory of Wolfram, for the 2-state, 3-site CA, it has $2^8$ rules. The Wolfram rules are defined from 0 to 255. Among the rules, eight Wolfram rules are linear. They are 0, 60, 90, 102, 150, 170, 204, and 240, respectively. The eight Wolfram linear rules are represented as follows:

\begin{equation}
 \begin{split}
&\texttt{Rule}\ 0:s_{i}(t+1)=0,\\
&\texttt{Rule}\ 60:s_{i}(t+1) = s_{i-1}(t)\oplus s_{i}(t),\\
&\texttt{Rule}\ 90:s_{i}(t+1) = s_{i-1}(t)\oplus s_{i+1}(t),\\
&\texttt{Rule}\ 102:s_{i}(t+1) = s_{i}(t)\oplus s_{i+1}(t),\\
&\texttt{Rule}\ 150:s_{i}(t+1) = s_{i-1}(t)\oplus s_{i}(t)\oplus s_{i+1}(t),\\
&\texttt{Rule}\ 170:s_{i}(t+1) = s_{i-1}(t),\\
&\texttt{Rule}\ 204:s_{i}(t+1) = s_{i}(t),\\
&\texttt{Rule}\ 240:s_{i}(t+1) = s_{t-1}(t).\\
 \end{split}
\end{equation}

Non-uniform CA is a special case of CA where not all cells present the same set of rules and these rules can change and evolve during the time.

The Wolfram rules 0, 60, 102, 170, 204 and 240 cannot produce high quality pseudorandom sequence, due to the combination of these rules cannot generate maximum-length sequences (m-sequences). However, the Wolfram rules 90 and 150 can generate m-sequences. The m-sequence can be generated by a transition matrix, and the transition matrix ($T$) is written as

\begin{equation}
T(x_n; y_n; z_n) = \left[
 \begin{matrix}
   y_1       & z_1   & 0       & \cdots\  & \cdots\ & 0        & 0\\
   x_2       & y_2   & z_2       &    \ddots       &        &         & 0\\
   0         & x_3   & y_3     &    \ddots       &  \ddots       &         & \vdots\\
    \vdots   &  \ddots    &  \ddots       &   \ddots        &  \ddots       &  \ddots        & \vdots\\
    \vdots   &     &  \ddots       &          & y_{n-2}       & z_{n-2}        & 0\\
   0         &     &        &   \ddots        & x_{n-1}       & y_{n-1}    & z_{n-1}\\
   0         & 0     & \cdots\ & \cdots\  & 0       & x_{n}    & y_{n}\\
  \end{matrix}
  \right]
\end{equation}

The transition matrix ($T$) can be re-written according to the rules 90 and 150 as

\begin{equation}
T(d_1, d_2, ..., d_{n-1}, d_n) = \left[
 \begin{matrix}
   d_1       & 1   & 0       & \cdots\  & \cdots\ & 0        & 0\\
   1       & d_2   & 1       &    \ddots       &        &         & 0\\
   0         & 1   & d_3     &    \ddots       &  \ddots       &         & 0\\
    \vdots   &  \ddots    &  \ddots       &   \ddots        &  \ddots       &  \ddots        & \vdots\\
    \vdots   &     &  \ddots       &          & d_{n-2}       & 1        & 0\\
   0         &     &        &   \ddots        & 1       & d_{n-1}    & 1\\
   0         & 0     & \cdots\ & \cdots\  & 0       & 1    & d_{n}\\
  \end{matrix}
  \right]
\end{equation}

Each element of the diagonal vector signifies linear rule according to

\begin{equation}
 d_i= \left\{
           \begin{array}{lr}
             0,\ \texttt{rule}\  90 &  \\
             1,\ \texttt{rule}\  150 &
             \end{array}
\right.
\end{equation}

 For a 8-bit hybrid rules 90 and 150 linear CA whose characteristic polynomial $f(x)$ can be described by

\begin{small}
\begin{equation}
 f(x)= T + x\texttt{I} \left[
 \begin{matrix}
  d_1+x     & 1   & 0       & \cdots\  & \cdots\ & 0   & 0     & 0\\
   1       & d_2+x   & 1       &    \ddots       &        &     &     & 0\\
   0         & 1   & d_3+x     &    \ddots       &  \ddots       &     &     & \vdots\\
    \vdots   &  \ddots    &  \ddots       &   d_4+x       &  \ddots       & \ddots   &       & \vdots\\
    \vdots   &     &  \ddots       &          & d_5+x        & 1     & 0   & 0\\
   \vdots         &     &        &   \ddots        & 1       & d_6+x    & 1 & 0\\
   0         &     &        &   \        & \ddots        &  1  & d_7+x & 1\\
   0         & 0     & \vdots & \cdots\  &   \cdots\      & 0  & 1  & d_8+x\\
  \end{matrix}
  \right]
\end{equation}
\end{small}

\begin{figure}[htp]
\centering
\begin{minipage}{0.65\linewidth}
\centering
\includegraphics[width=\textwidth]{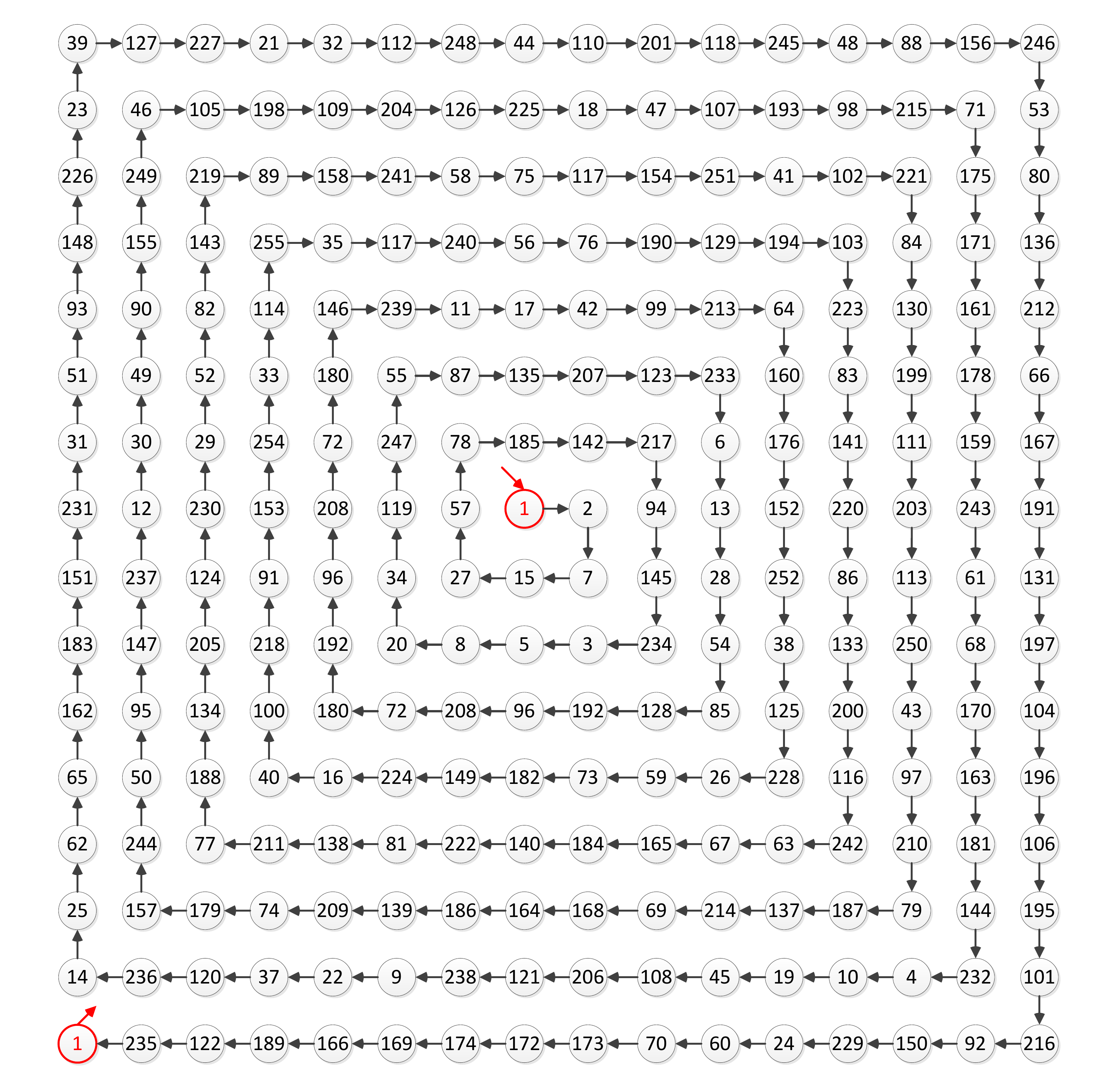}
(a)
\end{minipage}
\begin{minipage}{0.65\linewidth}
\centering
\includegraphics[width=\textwidth]{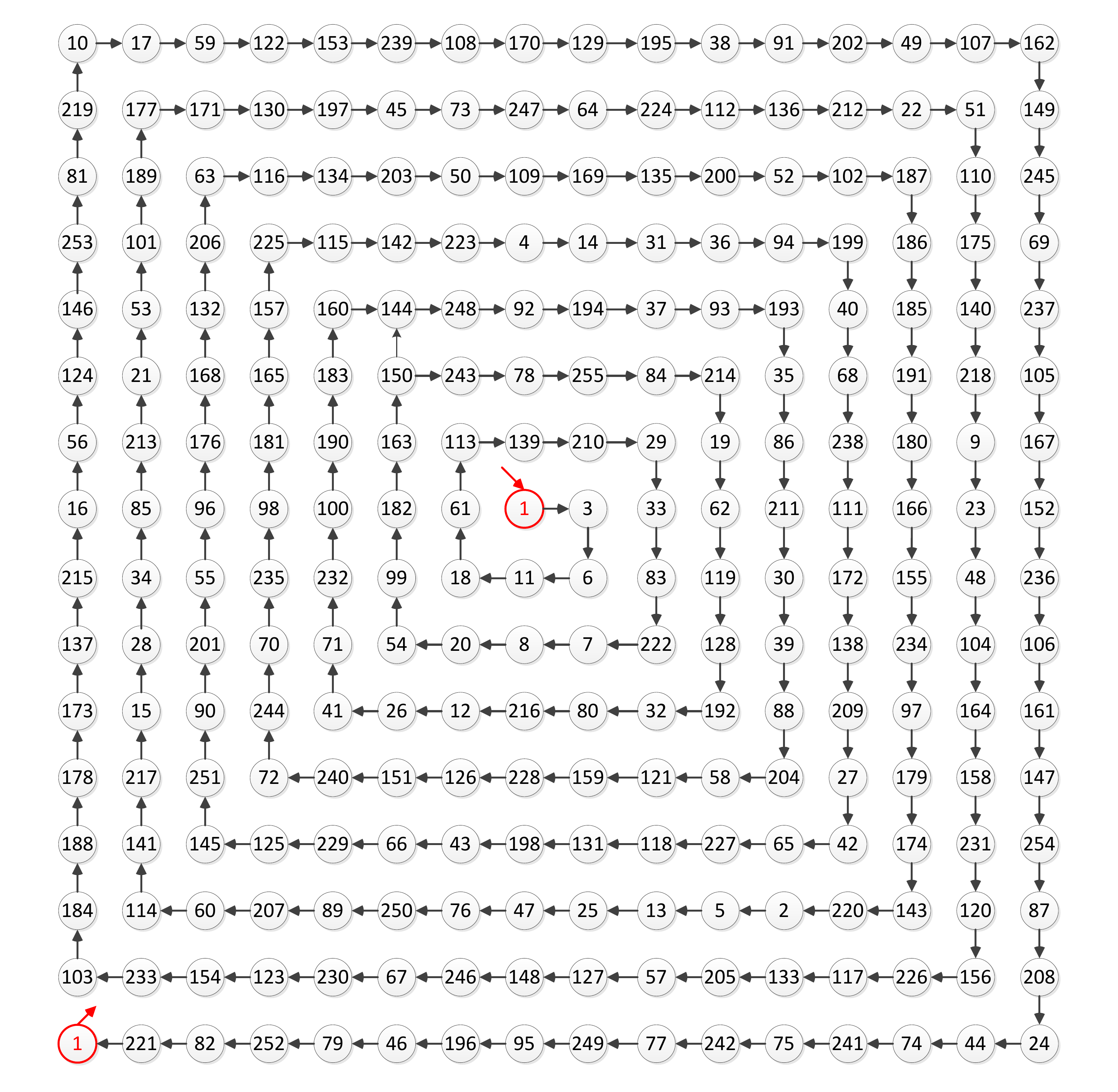}
(b)
\end{minipage}
\caption{The m-sequences generated by two hybrid rules 90 and 150:
(a) rules (150, 150, 90, 150, 90, 150, 90, 150), (b) rules (150, 90, 150, 90, 90, 90, 150, 90).}
\label{fig:Comparison}
\end{figure}

Let 8-bit group CA rules be $(150, 90, 150, 90, 90, 90, 150, 90)$, and $d_n = $(1, 0, 1, 0, $0,$ 0, $1, 0).$ The corresponding characteristic polynomial can be calculated as $f(x)=x^8+x^7+x^5+x^3+1$. The generated m-sequences with group CA rules $(150, 150, 90, 150, 90, 150, 90, 150)$ and $(150, 90, 150, 90, 90, 90, 150, 90)$ are shown in Fig.~\ref{fig:Comparison}.

\subsubsection{Chaotic sequences generated by the hyper-chaotic system}

\begin{figure}[!htp]
\centering
%\begin{minipage}[t]{\figwidth}
\centering
\includegraphics[width=\textwidth]{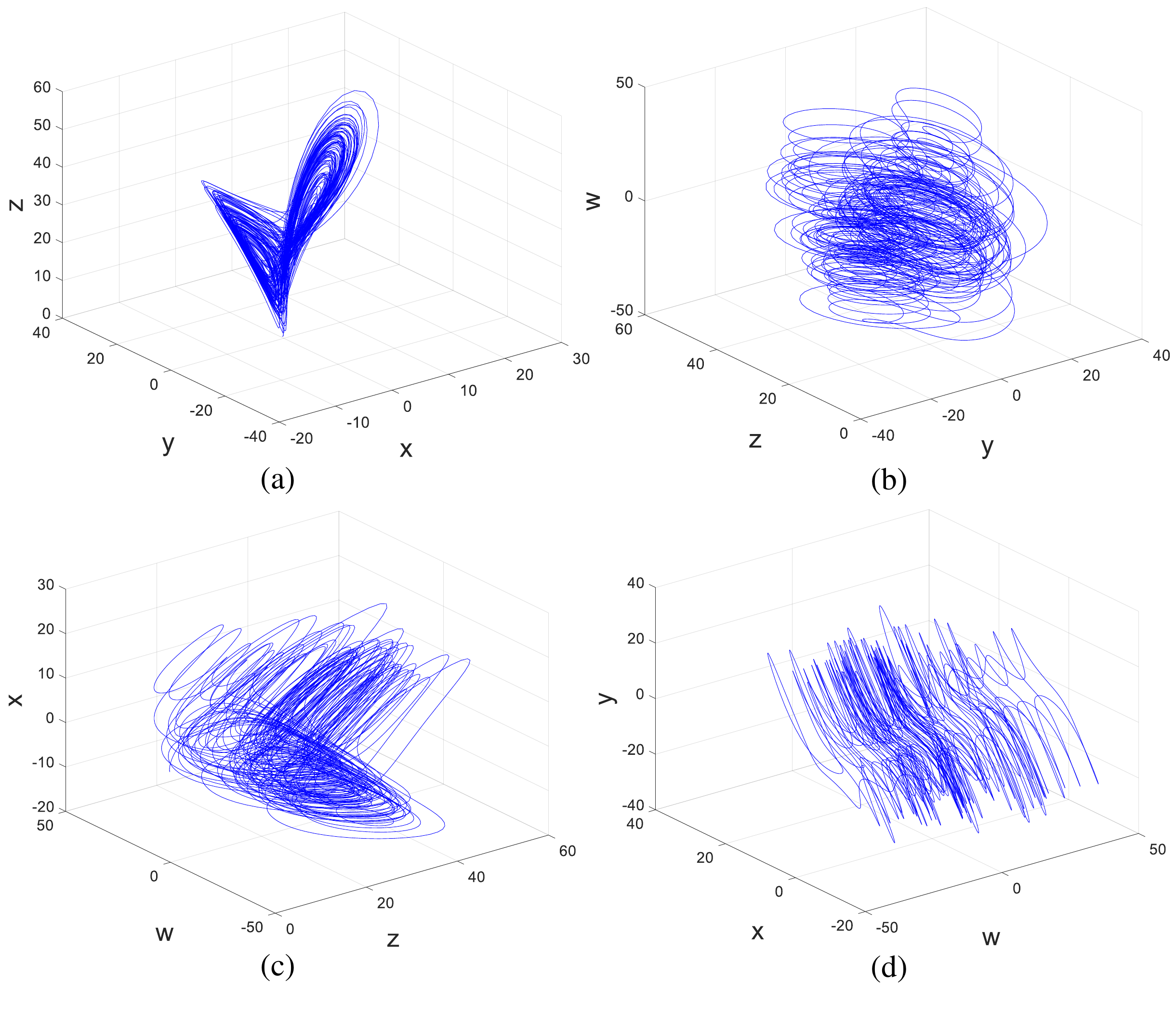}
%\end{minipage}
\caption{The attractors of the hyper-chaotic system: (a) x-y plane, (b) y-z plane, (c) z-w plane, and (d) w-x plane.}
\label{fig:chaos}
\end{figure}

The hyperchaotic system \cite{Diaconu2016,wang2008,xia2012,zhang2014,Ramasubramanian2000} is adopted to encrypt the captured monospectral EIs in our work. The hyperchaotic system is employed in this work  with four initial values $x_0$, $y_0$, $z_0$ and $\omega_0$:

\begin{equation}
\left\{
           \begin{array}{lr}
             \dot x = a(y-x)+ \omega &  \\
             \dot y = cx-y-xz  & \\
             \dot z = xy-bz  & \\
             \dot \omega = -yz+rw  & \\
             \end{array}
\right.
\end{equation}
where $a$, $b$, $c$ and $r$ denotes control parameters. Wang at el.~\cite{wang2008} have verified the dynamic features of the hyperchaotic system when $-6.43 < r < 0.17$. The Lorenz system behaves when $a$ = 10, $b$ = 8/3, $c$ = 28 and $11.52 < r < -0.06$, according to the method presented by Ramasubramanian et al. \cite{Ramasubramanian2000}.  We can obtain the Lyapunov exponents when $r = -1: \lambda_1 = 0.3381$, $\lambda_2 = 0.1586$, $\lambda_3 = 0$ and $\lambda_4 = -15.1752$. Fig.~\ref{fig:chaos} shows the attractors of hyperchaotic system with four different planes.

\subsubsection{Description of the encryption process}

Generally, for the previous color image encryption methods, three channels of the color image need to be extracted and parallel encrypted, respectively. It will increase extra time assume. However, in our proposed method, the captured monospectral EIs belongs to grayscale images, the encoding algorithm can directly encrypt them. It will greatly decrease the encryption calculation load. We assume the dimensional of the captured monospectral EIs is $L=M\times N$. The hyperchaotic system is used to permutate the pixel positions and m-sequence of CA is used to diffuse the pixel values. The schematic of the proposed encryption scheme has been described in Fig.~\ref{fig:en_sys}. The color image permutation and diffusion processes can be described as follows:

\begin{enumerate}[Step 1]
\item A color scene is captured by heterogeneous monospectral cameras, and the the monospectral EIs are obtained.

\item The pixel values of the monospectral EIs are reshaped as one dimensional vector $P = \{p_1, p_2, ..., p_{M\times N}\}$.

\item Generate four chaotic sequences $x_n = (x_1, x_2 , ..., x_{N_0+L/4})$, $y_n$ = $(y_1, y_2 ,$ $..., y_{N_0+L/4})$, $z_n = (z_1, z_2 , ..., z_{N_0+L/4})$, $w_n = (w_1, w_2 , ..., w_{N_0+L/4})$, by using Lorenz hyper-chaotic system under the condition that initial values are $x_0, y_0, z_0,$ and $w_0$. The system control parameters are set to $a, b, c, r$, where $a, b, c,$ and $r$, which can be gained by subsection 3.1. The size of each sequence is set to $N_0+L/4$. Generally, to remove the transient effect, the $N_0$ numbers of each of sequences is deleted. Thereby, the length of each of sequences is adjusted to $L/4$.

\item Combine four sequences $(x_n, y_n, z_n, w_n)$ with a new sequence $S_i$, and the length of the new sequence is $L$.

\item Sort the new hyperchaos sequence $S_i$ in ascending order, and then we obtain the index order sequence $I_S=\{I_{S_1}, I_{S_2}, ..., I_{S_L}\}$.

\item Rearrange the image gray value sequence $P$ by $I_S$ using the following equation, and then we can obtain the shuffled value $P'$:
\begin{equation*}
P_j'=P_{I_{S_j}}, \ j=1, ..., L
\end{equation*}
\end{enumerate}

In the previous permutation process, the monospectral EIs can achieve a perfect shuffled image utilize the above-mentioned 2D hyper-chaos system, but the pixel value of the original image cannot be changed. It is mean that this method is weak for statistical attacks. CA-based encoding algorithm can change the pixel value, and it provides different histogram distribution of the encrypted image. The color image diffusion process of our method is described as follow:
\begin{enumerate}[Step 1]
\item Two groups of 8-bit CA are used to generate two high quality pseudorandom sequences $f_x$ and $f_y$ (m-sequences) and $x, y \in (1, 2, ..., 255)$, the m-sequence generation algorithm can be found in Subsection~\ref{subsubsection:2.21}.

\item Let $f_x$ be the main body generator, we randomly select $k$ integers from m-sequence $f_y$. Here, $k= \left \lceil \frac{L}{255}\right\rceil$, the selected integer $k$ as the seed, we can generate $k$ different m-sequences $f_{y(i)}$, and $i \in (1, 2, ..., k)$.

\item Each of the new generated m-sequences do the exclusive or operation with the seed sequence $f_x$ according to
\begin{equation*}
Z_i= f_x\oplus f_y(i),\ k =1, ..., k
\end{equation*}

\item According to the step 3, we can generate $k$ sequences $Z_i$, these sequences can combined to one sequence $Z'$ according to different arrangements. The length of the sequence $Z'$ is $255\times k$ and greater than or equal to the length of the shuffled image sequence $L$. When we do the exclusive or operation with the shuffled image sequence $L$, the length of the sequence $Z'$ should be adjusted to $L$.

\item The cipher image $C$ of the proposed method can be generated by

\begin{equation*}
C=P'\oplus(f_x\oplus f_y(i))
\end{equation*}

\end{enumerate}

\subsubsection{Description of the decryption process:}

Color image decryption process is the inverse of the color image encryption process. For convenience, the color image decryption process of our method is simply illustrated by the following steps.

\begin{enumerate}[Step 1]

\item According to the provided linear two groups CA rules, two m-sequences can be generated to decode the cipher image.

\item By utilizing the previous step, we can obtain the shuffled image, and the monospectral EIs can be decrypted by using the inverse encoding process of the hyper-chaotic system.

\item By using the improved SR reconstruction algorithm, which has discussed in Sec., the color image can be reconstructed from the decrypted monospectral EIs.

\end{enumerate}

\subsection{Color image reconstruction by improved SR reconstruction algorithm}

\begin{figure}[!htb]
\centering
%\begin{minipage}[t]{\figwidth}
\centering
\includegraphics[width=\textwidth]{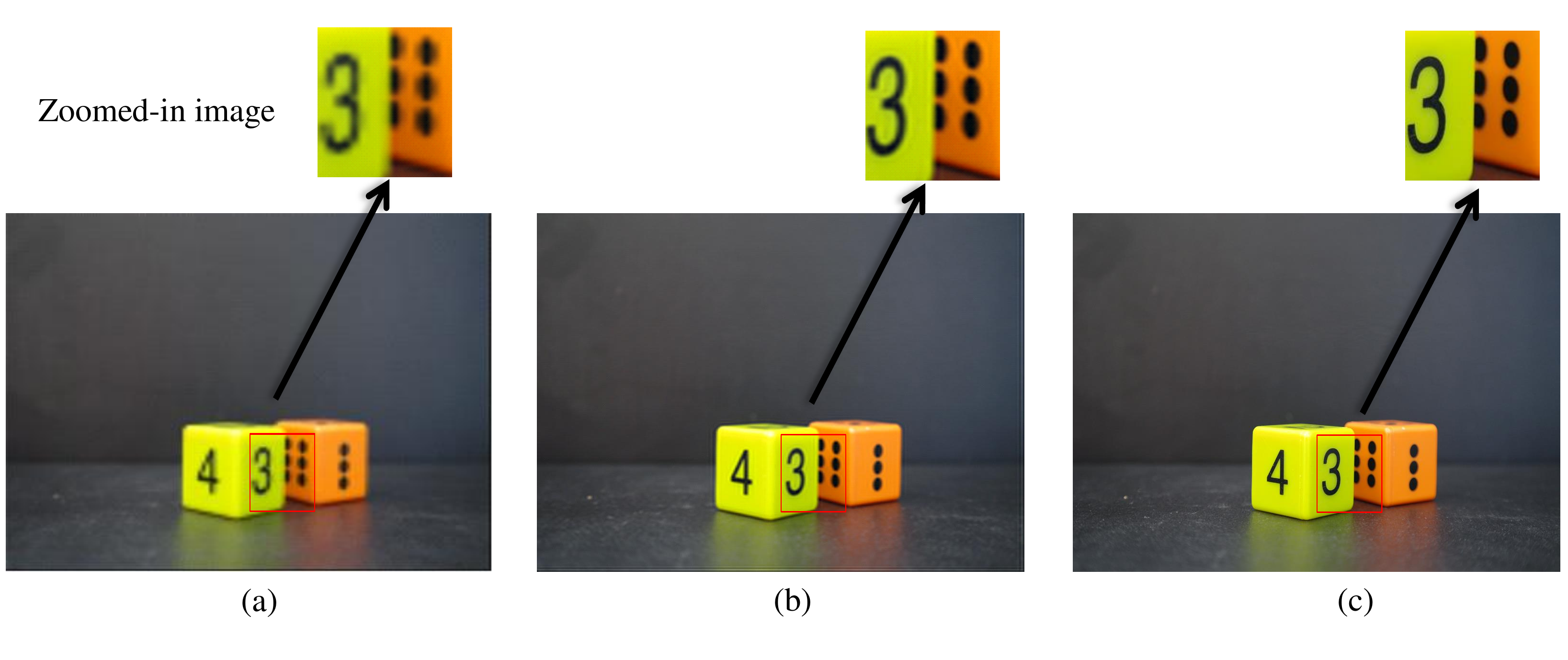}
%\end{minipage}
\caption{Simulation results with three kinds of resolution enhancement methods: (a) CIIR method (pickup distance $l$ = 12 mm), (b) $L_2$ + bilateral TV, (c) proposed method.}
\label{fig:reconstruct}
\end{figure}

The conventional CIIR method magnifies each pixels of the elemental image according to the magnified factor before superimposing onto the output plane. The overlapped process and superimposed process of the adjacent pixels from the elemental images will result in blur noise, and thereby reduces the image quality, as shown in Fig.~\ref{fig:reconstruct} (a). Therefore, in order to realize the high quality image, we study an adaptive and widely used multichannel SR and image restoration algorithm \cite{Bose2015,Ji2015,Gunturk2004,wang2017,park2003,Farsiu2006,Schroder2010}. For color image reconstruction, because the reconstruction method of three color channels is identical, for clarity of the discussion, we will restrict our presentation to estimate a single color channel (green channel). We can mathematically model the multichannel imaging system. Let $x$ be the desired high-resolution image and $y_k$  be the low-resolution images captured by the $k-th$ green color sensor. The goal is to estimate $x$ from low-resolution observations $( y_k)$. A measurement from the multichannel imaging system consists of an array of sub-image and can be modeled as

\begin{equation}
y_k = D_kH_kW_kx + n_k,  k = 0, 1, 2, 3, ..., N,
\end{equation}
where $N$ represents the total number of the monospectral sensor (Red, Green or Blue). $x$ (of size $[\gamma^2M^2\times1]$) and $y_k$ (of size $M^2\times 1$) denote the unknown high-resolution image and the captured $k-th$ blurred, low-resolution and noisy sub-image. $\gamma$ denotes the upsampling factor. The matrix $D_k$ (of size $[M^2\times \gamma^2M^2]$) and matrix $H_k$ (of size $[\gamma^2M^2 \times \gamma^2M^2]$) denote the spatial subsampling by the sensor and the point-spread function, respectively. $W_k$ is the warping matrix (of size $[\gamma^2M^2 \times \gamma^2M^2]$) that represents the total displacement (geometric distortion and parallax shift) of the $k-th$ image seen from the $k-th$ sensor. $n_k$ is the image noise term associated with the $k-th$ sub-image. Thus, the maximum likelihood estimation of the high-resolution image can be calculated by

\begin{equation}
\hat{x} = \mathop{\arg\min}_{x}\left[\sum_{k=1}^{M}\rho(D_kH_kW_k,y_k)\right],
\end{equation}
or by an implicit equation

\begin{equation}
\sum_{k}\Psi(D_kH_kW_k,y_k)=0,
\end{equation}
where $\Psi(D_kH_kW_k,y_k)=(\partial/\partial x)(D_kH_kW_k-y_k)$  and $\rho$ represents the ``distance'' between the measurements and model. To get the estimate result of image $\hat{x}$, the $\rho$ can be set to $L_2$ norm

\begin{equation}
\hat{x} = \mathop{\arg\min}_{x}\left[\sum_{k=1}^{M}\|D_kH_kW_k-y_k\|_{2}^2\right]
\end{equation}

Due to SR reconstruction algorithm is an ill-posed problem, the square and over-determined solutions are not stable. The small amounts of attack will result in huge perturbation. To address this problem, we add a regularization term to resolve the ill-posed problem.

\begin{equation}
\hat{x} = \mathop{\arg\min}_{x}\left[\|\sum_{k=1}^{M}\rho(D_kH_kW_k,y_k)\|_{2}^2+\lambda\Gamma(x)\right]
\label{eq:hatx}
\end{equation}
where $\lambda$ represents a regularization parameter and $\Gamma$ denotes the regularization cost function. The most widely used cost regularization function is the bilateral total variation (TV) cost function \cite{wang2017,park2003}. $\Gamma(x)=\|\Upsilon x\|_1$, where $\Upsilon$ is a high pass operator such as Laplacian, derivative, or even identity matrix.

To determine the solution for Eq.~(\ref{eq:hatx}), we utilize steepest descent to optimize the solution, the iterative update for $(\hat{x})$ is written as

\begin{equation}
\hat{x}^{(n+1)} = \hat{x}^{(n)}-\beta\{W_k^TH_k^TD_k^T sign(D_kH_kW_k-y_k)+\lambda\Gamma(x)\},
\label{eq:hatxx}
\end{equation}
where $\beta$ denotes a scalar factor. In order to import robustness into this procedure, the term $W_k^TH_k^TD_k^T sign(\cdot)$ of Eq.~(\ref{eq:hatxx}) can be replaced with a scaled pixel-wise median adaptive filter:

\begin{equation}
\Delta_{k} \triangleq n\cdot median\{W_k^TH_k^TD_k^T sign(D_kH_kW_k-y_k)\}_{k=1}^n
\end{equation}

We choose to use the median pixel-wise estimator method, and the update equation that minimizes Eq.~(\ref{eq:hatxx}) is written as

\begin{equation}
\hat{x}^{(n+1)} = \hat{x}^{(n)}-\beta\{\Delta_{k}+\lambda\Gamma(x)\},
\end{equation}

The iteration is terminated when the calculated error $e^{(n)}$ is less than the given threshold value $\eta$. Otherwise, the inferred SR reconstructed image is updated by Eq.~(\ref{eq:hatxx}). The error function $e^{(n)}$ is described by

\begin{equation}
e^{(n)}=\frac{\|\hat{x}^{(n+1)}-\hat{x}^{(n)}\|_2}{\|\hat{x}^{(n)}\|_2}
\end{equation}

We have reconstructed a single color channel, namely green channel by the above mentioned SR method, and the other two color channels also easily estimated by the same way. The high-resolution color image can be fused by these three color channels and the fused color image is shown in Fig.~\ref{fig:reconstruct} (c). SR reconstruction method which is based on $L_2$ regularization and bilateral TV cost function resulted in Fig.~\ref{fig:reconstruct} (b). It is clear that the SR reconstruction method ($L_2$ + bilateral TV) is far better than the CIIR approach (Fig.~\ref{fig:reconstruct} (a)), but appearing some artifacts. From the simulation results, it is clear that the proposed method effectively improves image-quality of the reconstructed image.

\section{Simulation results and performance analysis}
\label{sec:3}

We now confirm the performance of the proposed method through simulation simulations. In our experiment, the monospectral image imaging setup consists of $6 \times 6$ low cost monospectral cameras. The focal length of each camera is set to 2.8 mm and the distance between two adjacent cameras is 9 mm. To decrease the cost of camera consume, the synthetic-aperture integral imaging pickup system is used, which composed of three monospectral (R, G, and B) cameras and a motorized translation stage. We utilize each camera to capture three-group sub-image arrays, and obtain 108 sub-images in total, here, we choose 36 different monospectral sub-images (concluding R, G, and B spectrums) from them as the monospectral SIA to reconstruct the high-resolution image. Each camera provides $800 \times 600$ pixels. For CA encryption, we use two-group linear CA rules for image encryption, they are rules $(150, 150, 90, 150, 90, 150, 90, 150)$ and rules $(150, 90, 150, 90, 90, 90, 150, 90)$, respectively. The integer $N_0$ of the hyper-chaotic system is set to 150.

\subsection{Key sensitive analysis}

\begin{figure}[!htp]
\centering
%\begin{minipage}[t]{\figwidth}
\centering
\includegraphics[width=\textwidth]{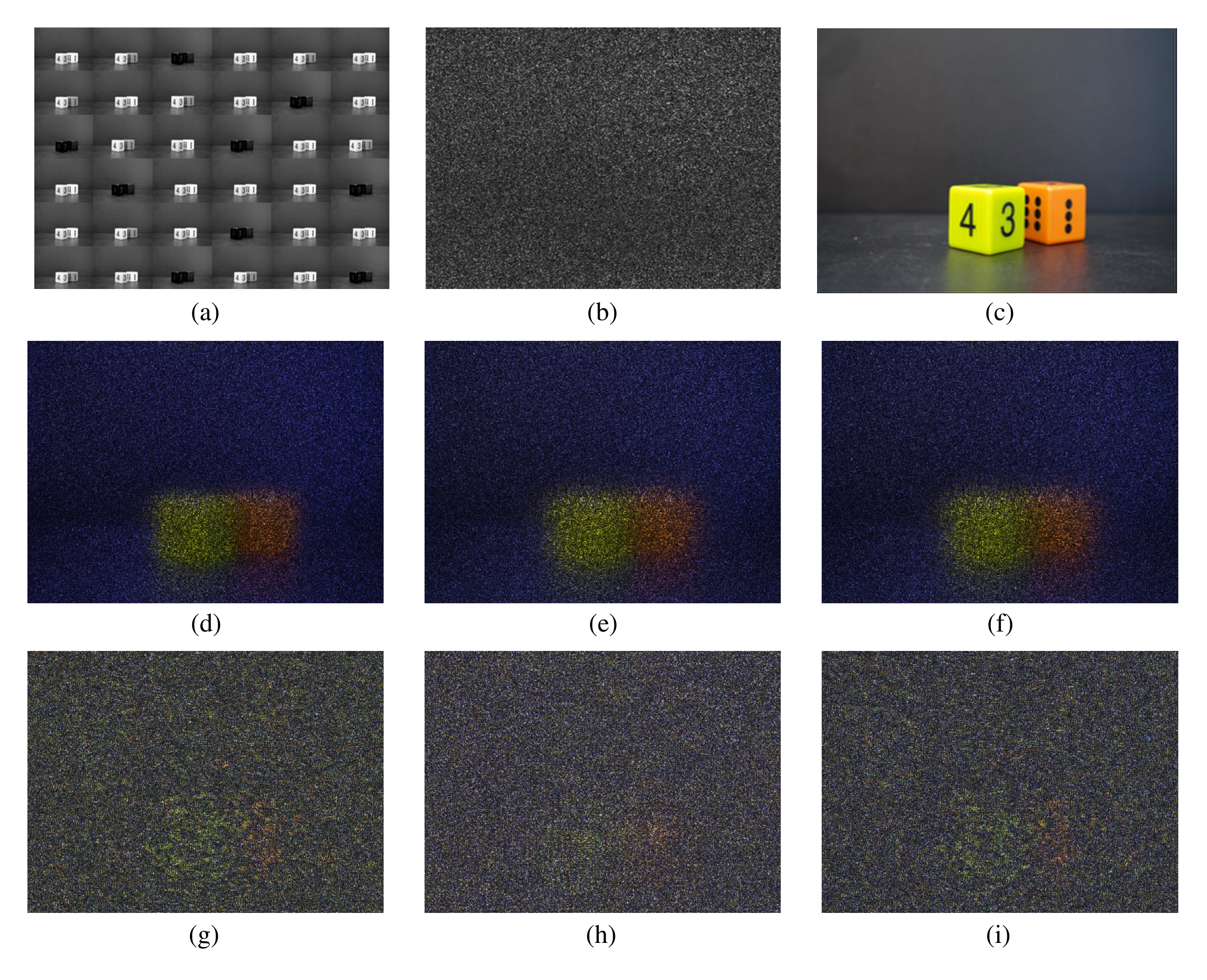}
%\end{minipage}
\caption{(a) Captured SIA by the monospectral camera array, (b) encrypted image, (c) reconstructed color image with all correct keys, (d)-(e) decrypted images with CA rule changed: (d) one bit changed, (e) two bits changed, (f) three bits changed, (g) initial values $x_0$ is wrong, (h) initial values $y_0$ is wrong, (i) initial values $z_0$ is wrong.}
\label{fig:encryption}
\end{figure}

A good encryption method requires high key sensitivity. Fig.~\ref{fig:encryption} (a) shows the captured monospectral SIA by the monospectral camera array. Fig.~\ref{fig:encryption} (b) shows the encrypted image by the proposed encryption method. Fig.~\ref{fig:encryption} (c) shows the reconstructed color image. Figs.~\ref{fig:encryption} (d)-(f) show the recovered color images with wrong CA rules. Figs.~\ref{fig:encryption} (g)-(i) show the recovered color images with wrong input initial $x_0$, $y_0$, $z_0$. he results show that decryption can only be successfully completed using all encryption keys. Meantime, the proposed method has a large secret key space, which includes not only the initial conditions and hyper-chaotic system parameters $x_0$, $y_0$, $z_0$, $w_0$, and $N_0$, but also the additional key space provided by CA.

\subsection{Statistical analysis}

\begin{figure}[htbp]
\centering
%\begin{minipage}[t]{\figwidth}
\centering
\includegraphics[width=\textwidth]{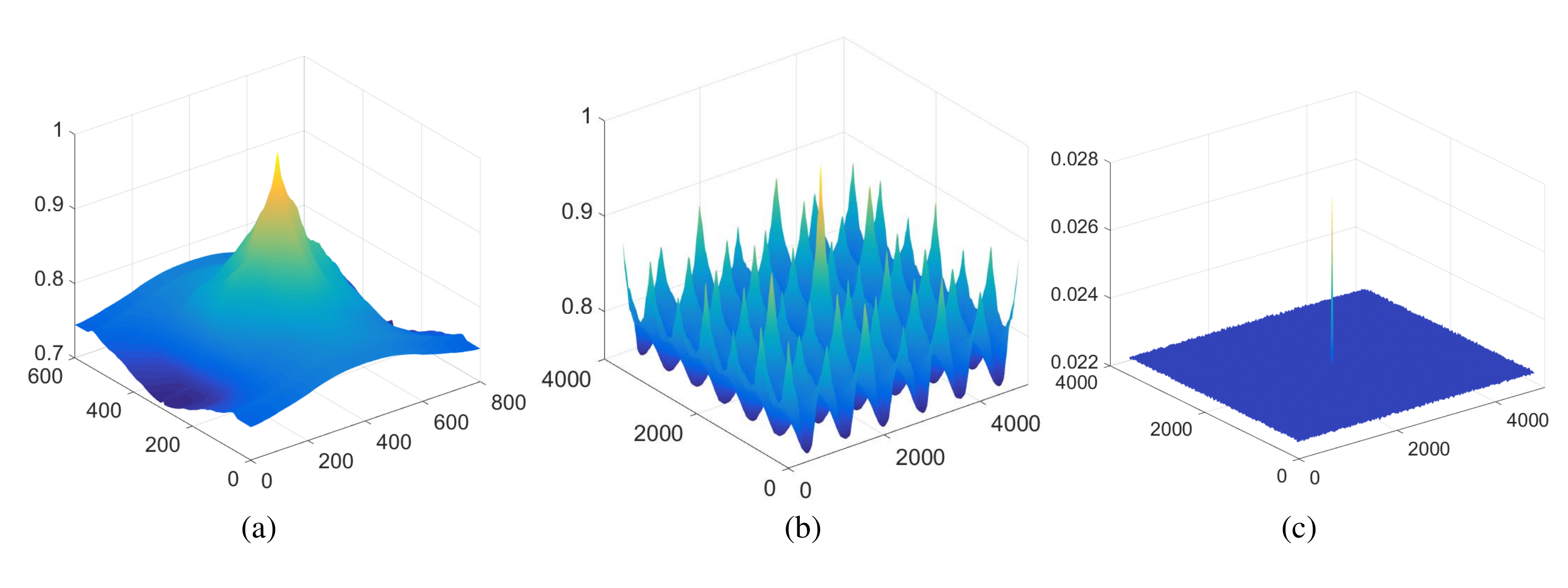}
%\end{minipage}
\caption{(a) Autocorrelation of one sub-image, (b) autocorrelation of the captured SIA, (c) autocorrelation of SIA after encryption.}
\label{fig:auto}
\end{figure}

There is no statistical similarity in the encrypted image which is a very important key to prevent data leakage. Thereby, an idea encryption method should provide nearly no correlation in the adjacent pixels of the encrypted image. The sub-image, as previously mentioned, only contains a single spectral of color space. Therefore, the SIA resembles gray-scale image and each sub-image only captures one kind of color channel information at every pixel location. Thus the encrypted SIA also is a gray-scale image. Figs.~\ref{fig:auto} (a) and (b) show autocorrelation of one sub-image and SIA. Fig.~\ref{fig:auto} (c) exhibits autocorrelation of the encrypted image; whereas the autocorrelation of the encrypted image is much weakness than that of the original sub-image. Hence, the proposed encryption method provides high robust against the statistical attacks.

\begin{figure}[htbp]
\centering
%\begin{minipage}[t]{\figwidth}
\centering
\includegraphics[width=\textwidth]{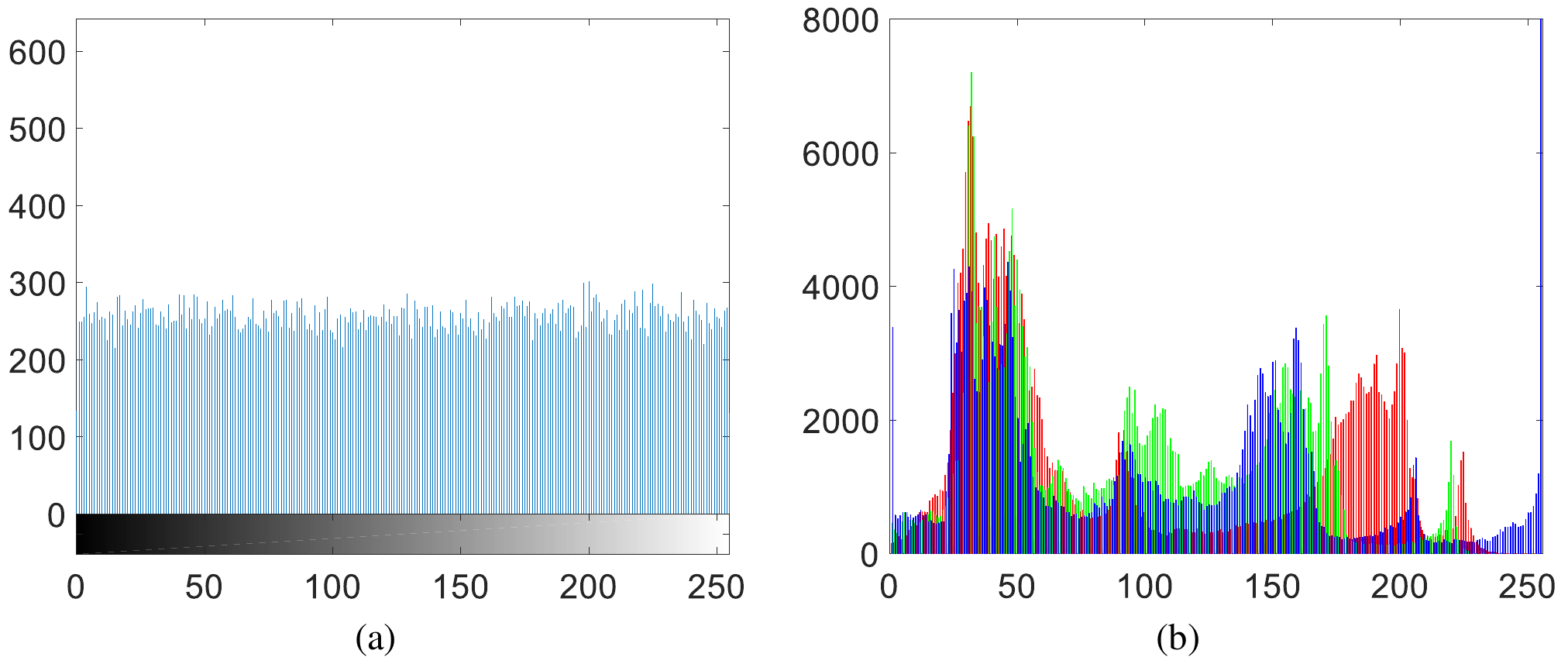}
%\end{minipage}
\caption{(a) Histogram of the encrypted image, (b) histogram of the decrypted color image.}
\label{fig:histogram}
\end{figure}

We also analyze the pixel distribution of the SIA after encoding and reconstruction. Fig.~\ref{fig:histogram} (a) shows the histogram of the histogram of SIA after encoding and Fig.~\ref{fig:histogram} (b) shows the reconstructed color image by the decrypted SIA. On the other hand, we can also say that the pixel distribution of the input sub-image is quite different from that of the sub-image after encoding.

\subsection{Robustness analysis}

\begin{figure}[htbp]
\centering
%\begin{minipage}[t]{\figwidth}
\centering
\includegraphics[width=0.9\textwidth]{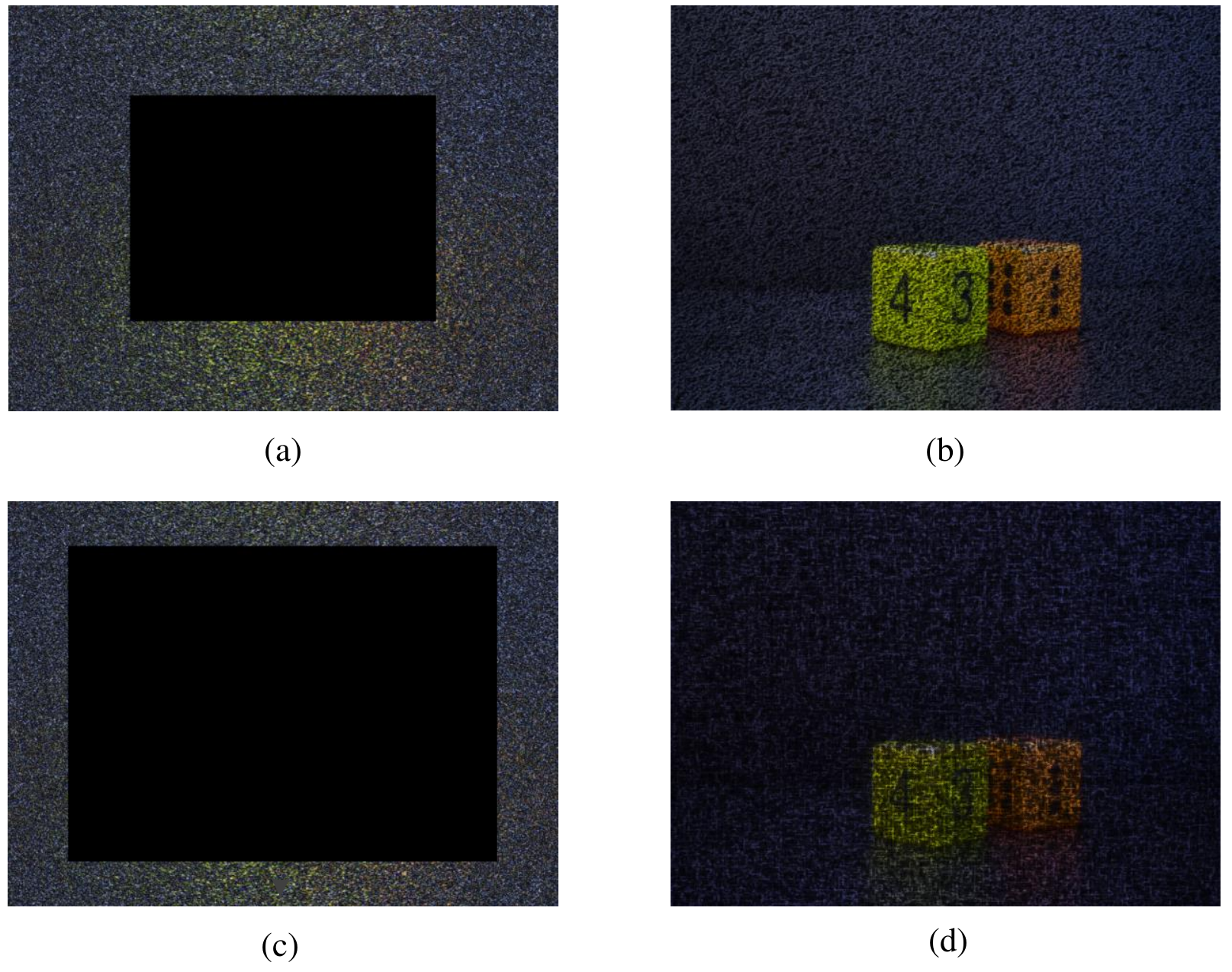}
%\end{minipage}
\caption{Decrypted color images against occlusion attack: (a) 30\% pixels of encrypted image were occluded, (b) the image decrypted from (a), (c) 70\% pixels of encrypted image were occluded, (d) the decrypted image from (c).}
\label{fig:occluded}
\end{figure}

An idea image encryption algorithm should resist attacks. Fig.~\ref{fig:occluded} (a) shows the encrypted image against occlusion attack when 30\% pixels were occluded. Fig.~\ref{fig:occluded} (b) shows the decrypted image from Fig.~\ref{fig:occluded} (a). Fig.~\ref{fig:occluded} (c) shows the attacked encrypted image when 70\% pixels were occluded and Fig.~\ref{fig:occluded} (d) shows the corresponding decrypted image. The experimental results show that even though 70\% pixels were occluded, the image can be reconstructed successfully. In other words, our method provides high robustness when the encrypted images against data loss attack.

\subsection{Encryption calculation analysis}

Besides to security and robustness consideration, the calculation speed is also important for encryption method, especially for real-time Internet applications. To verify the effectiveness of our work, numerical simulations are tested under the environment of software Matlab 8.1. Table \ref{tab:1} shows the test results applied to color images with different sizes.

% For tables use
\begin{table}[!htp]
\centering
% table caption is above the table
\caption{The runtime performance test (Sec.)}
\label{tab:1}       % Give a unique label
% For LaTeX tables use
\begin{tabular}{lllll}
\hline\noalign{\smallskip}
%Color image size  && Encryption time    \\
Color image size & Ref.\cite{li2012} & Ref.\cite{zhou2016} & Ref.\cite{lix2016} & Proposed\\
\noalign{\smallskip}\hline\noalign{\smallskip}
800 $\times$ 600 & 0.51 & 0.63 & 0.73 & 0.16 \\
2400 $\times$ 1800 & 2.32 & 2.83 & 4.11 & 0.83 \\
4800 $\times$ 3600 & 8.13 & 8.89 & 9.13 & 3.06 \\
\noalign{\smallskip}\hline
\end{tabular}
\end{table}

\section{Conclusion}
\label{sec:4}

We have presented a color hyperchaotic image encryption method implemented by the combined use of the monospectral imaging technique and the CA encoding algorithm. We have utilized a monospectral camera array and an improved SR reconstruction for color image information acquisition and reconstruction. The simulation results confirmed that the proposed method has effectively eliminated color crosstalk and improved security by increasing the key space in the decoding process. We also use different security measures to verify the performance of the proposed method. The simulation results prove that the proposed image encryption method has great security and robustness.

\section*{Acknowledgement}
This work was supported by National Key R\&D Program of China (2017YFB10 02900), National Natural Science Foundation of China (NSFC) (61705146, 61535007), and Fundamental Research Funds for the Central Universities (YJ201637).

\section*{References}
\bibliography{References}
\begin{small}

\end{small}
%\label{sec:5}

%% The Appendices part is started with the command \appendix;
%% appendix sections are then done as normal sections
%% \appendix

%% \section{}
%% \label{}

%% If you have bibdatabase file and want bibtex to generate the
%% bibitems, please use
%%
%%  \bibliographystyle{elsarticle-num}
%%  \bibliography{<your bibdatabase>}

%% else use the following coding to input the bibitems directly in the
%% TeX file.

%\begin{thebibliography}{00}

%% \bibitem{label}
%% Text of bibliographic item

%\bibitem{}

%\end{thebibliography}
\end{document}